\documentclass{PoS}

\title{High Temperature Confinement in SU(N) Gauge Theories }

\ShortTitle{High Temperature Confinement in SU(N) Gauge Theories }

\author{\speaker{Michael Ogilvie}%
        \thanks{MCO is supported by a grant from the U.S. Dept. of Energy.}\\
       Washington University\\
       E-mail: \email{mco@physics.wustl.edu}}
\author{Peter N. Meisinger\\
        Washington University\\
        E-mail: \email{pnm@physics.wustl.edu}}

\abstract{$SU(N)$ gauge theories, extended with adjoint fermions having periodic
boundary conditions, are confining at high temperature for sufficiently
light fermion mass $m$. Lattice simulations indicate that this confining
region is smoothly connected to the confining region of low-temperature
pure $SU(N)$ gauge theory. In the high temperature confining region,
the one-loop effective potential for Polyakov loops has a $Z(N)$-symmetric
confining minimum. String tensions associated with Polyakov loops
are smooth functions of $m/T$. In the magnetic sector, the Polyakov
loop plays a role similar to a Higgs field, leading to a breaking
of $SU(N)$ to $U\left(1\right)^{N-1}$. This is turn yields an effective
theory where magnetic monopoles give rise to string tensions for spatial
Wilson loops. These string tensions are calculable semiclassically.
There are many analytical predictions for the high-temperature region
that can be tested by lattice simulations, but lattice work will be
crucial for exploring the crossover from this region to the low-temperature
confining behavior of pure gauge theories.}

\FullConference{The XXVI International Symposium on Lattice Field Theory \\
		 July 14 - 19, 2008\\
		 Williamsburg, Virginia, USA}

\begin{document}

\section{Introduction}

One of the long-standing problems of modern strong-interaction physics
is the origin of quark confinement. Finite temperature gauge theories
are advantageous in many aspects for the study of confinement. The
Polyakov loop operator $P$, given by\begin{equation}
P\left(\vec{x}\right)=\mathcal{P}\exp\left[i\int_{0}^{\beta}dtA_{4}\left(\vec{x},t\right)\right]\end{equation}
 represents the insertion of a static quark into a thermal system
of gauge fields. It is the order parameter for the deconfinement phase
transition in pure $SU(N)$ gauge theories, with $\left\langle Tr_{F}P\right\rangle =0$
in the confined phase, and $\left\langle Tr_{F}P\right\rangle\ne0$
in the deconfined phase. The deconfinement phase transition is associated
with the spontaneous breaking of a global $Z(N)$ symmetry $P\rightarrow zP$
where $z=\exp\left(2\pi i/N\right)$ is the generator of $Z(N)$. 

It is remarkable that there is a simple class of $Z(N)$-invariant
systems that are confining at arbitrarily high temperatures, evading
the transition to the deconfined phase found in the pure gauge theory.
A high-temperature confined phase is obtained from a pure gauge theory
by the addition of fermions in the adjoint representation of $SU\left(N\right)$,
with the non-standard choice of periodic boundary conditions
for the fermions in the
timelike direction. 
If the number of adjoint fermion flavors $N_{f}$ is not too large,
these systems are asymptotically free
at high temperature, and therefore
the effective potential for $P$ is calculable using perturbation theory.
The system
will lie in the confining phase
if the fermion mass $m$ is sufficiently light and $N_{f} > 1/2$.
In this case, electric string tensions can be calculated
perturbatively from the effective potential, and magnetic string tensions
arise semiclassically from non-Abelian magnetic monopoles. Thus the
high-temperature confining phase provides a realization of one of
the oldest ideas about the origin of confinement.

\section{High Temperature Confinement}

The one-loop effective potential for a boson in a representation $R$
with spin degeneracy $s$ moving in a Polyakov loop background $P$ at
non-zero temperature and density is given by \cite{Gross:1980br,Weiss:1980rj}
\begin{equation}
V_{b}=sT\int\frac{d^{d}k}{\left(2\pi\right)^{d}}Tr_{R}\left[\ln\left(1-Pe^{\beta\mu-\beta\omega_{k}}\right)+\ln\left(1-P^{+}e^{-\beta\mu-\beta\omega_{k}}\right)\right].\end{equation}
Periodic boundary conditions are assumed. With standard boundary conditions
(periodic for bosons, antiperiodic for fermions), 1-loop effects always
favor the deconfined phase. For the case of pure gauge theories, the
one-loop effective potential can be written in the form

\begin{equation}
V_{gauge}\left(P,\beta,m,N_{f}\right)=\frac{-2}{\pi^{2}\beta^{4}}\sum_{n=1}^{\infty}\frac{Tr_{A}P^{n}}{n^{4}}.\end{equation}
This series is minimized, term by term if $P\in Z(N)$, so $Z(N)$
symmetry is spontaneously broken at high temperature. The same result
is obtained for any bosonic field with periodic boundary conditions
or for fermions with antiperiodic boundary conditions. 

The addition of fermions with periodic boundary conditions can restore
the broken $Z(N)$ symmetry. Consider the case of $N_{f}$ flavors
of Dirac fermions in the adjoint representation of $SU(N)$. Periodic
boundary conditions in the timelike direction imply that the generating
function of the ensemble, \emph{i.e.}, the partition function, is
given by\begin{equation}
Z=Tr\left[\left(-1\right)^{F}e^{-\beta H}\right]\end{equation}
where $F$ is the fermion number. This ensemble, familiar from supersymmetry,
can be obtained from an ensemble at chemical potential $\mu$ by the
replacement $\beta\mu\rightarrow i\pi$. In perturbation theory, this
shifts the Matsubara frequencies from $\beta\omega_{n}=\left(2n+1\right)\pi$
to $\beta\omega_{n}=2n\pi$. The one loop effective potential is like
that of a bosonic field, but with an overall negative sign due to
fermi statistics \cite{Meisinger:2001fi}. The sum of the effective potential for the
fermions plus that of the gauge bosons gives\begin{equation}
V_{1-loop}\left(P,\beta,m,N_{f}\right)=\frac{1}{\pi^{2}\beta^{4}}\sum_{n=1}^{\infty}\frac{Tr_{A}P^{n}}{n^{2}}\left[2N_{f}\beta^{2}m^{2}K_{2}\left(n\beta m\right)-\frac{2}{n^{2}}\right].\end{equation}
Note that the first term in brackets, due to the fermions, is positive
for every value of $n$, while the second term, due to the gauge bosons,
is negative. 

The largest contribution to the effective potential at high temperatures
is typically from the $n=1$ term, which can be written simply as\begin{equation}
\frac{1}{\pi^{2}\beta^{4}}\left[2N_{f}\beta^{2}m^{2}K_{2}\left(\beta m\right)-2\right]\left[\left|Tr_{F}P\right|^{2}-1\right]\end{equation}
 where the overall sign depends only on $N_{f}$ and $\beta m$. If
$N_{f}\ge1$ and $\beta m$ is sufficiently small, this term will
favor $Tr_{F}P=0$. On the other hand, if $\beta m$ is sufficiently
large, a value of $P$ from the center, $Z(N)$, is preferred. Note
that an $\mathcal{N}=1$ super Yang-Mills theory would correspond to $N_{f}=1/2$ and
$m=0,$ giving a vanishing perturbative contribution for all $n$
\cite{Davies:1999uw,Davies:2000nw}.
This suggests that it should be possible to obtain a $Z(N)$ symmetric,
confining phase at high temperatures using adjoint fermions with periodic
boundary conditions or some equivalent deformation of the theory.

This possibility has been confirmed in $SU(3)$, where both lattice
simulations and perturbative calculations have been used to show that
a gauge theory action with an extra term of the form $\int d^{4}x\ a_{1}Tr_{A}P$
is confining for sufficiently large $a_{1}$ at arbitrarily high temperatures
\cite{Myers:2007vc}.
This simple, one-term deformation is sufficient for $SU(2)$ and
$SU(3)$. However, in the general case, a deformation with at least
$\left[\frac{N}{2}\right]$ terms is needed to assure confinement
for representations of all possible non-zero $k$-alities. Thus the
minimal deformation necessary is of the form
\begin{equation}
\sum_{k=1}^{\left[\frac{N}{2}\right]}a_{k}Tr_{A}P^{k}\end{equation}
which is analyzed in detail in Joyce Myer's presentation in these procedings \cite{Myers:2008ey}. If all the coefficients
$a_{k}$ are sufficiently large and positive, the free energy density\begin{equation}
V_{1-loop}\left(P,\beta,m,N_{f}\right)=\frac{-2}{\pi^{2}\beta^{4}}\sum_{n=1}^{\infty}\frac{Tr_{A}P^{n}}{n^{4}}+\sum_{k=1}^{\left[\frac{N}{2}\right]}a_{k}Tr_{A}P^{k}\end{equation}
will be minimized by a unique set of Polyakov loop eigenvalues corresponding
to exact $Z(N)$ symmetry. 

The unique set of $SU(N)$ Polyakov eigenvalues invariant under Z(N)
is $\left\{ w,wz,wz^{2},..,wz^{N-1}\right\} $, where $z=e^{2\pi i/N}$
is the generator of $Z(N)$, and $w$ is a phase necessary to ensure
unitarity \cite{Meisinger:2001cq}. A matrix with these eigenvalues, such as
$P_{0}=w\cdot diag\left[1,z,z^{2},..,z^{N-1}\right]$,
 is gauge-equivalent to itself after a $Z(N)$ symmetry operation: 
$zP_{0}=gP_{0}g^{+}$.
This guarantees that
 $Tr_{F}\left[P_{0}^{k}\right]=0$
for any value of $k$ not divisible by $N$,
indicating confinement for all representations transforming
non-trivially under $Z(N)$. 

To prove that $P_{0}$ is a global minimum of the effective potential,
we use the high-temperature expansion for the one-loop free energy
of a particle in an arbitrary background Polyakov loop gauge equivalent
to the matrix $P_{jk}=\delta_{jk}e^{i\phi_{j}}$.
The first two terms have the form \cite{Meisinger:2001fi}
\begin{eqnarray}
V_{1-loop}&\approx&\sum_{j,k=1}^{N}(1-\frac{1}{N}\delta_{jk})\frac{2\left(2N_{f}-1\right)T^{4}}{\pi^{2}}\left[\frac{\pi^{4}}{90}-\frac{1}{48\pi^{2}}\left(\phi_{j}-\phi_{k}\right)^{2}\left(\phi_{j}-\phi_{k}-2\pi\right)^{2}\right]\nonumber\\
&&-\sum_{j,k=1}^{N}(1-\frac{1}{N}\delta_{jk})\frac{N_{f}m^{2}T^{2}}{\pi^{2}}\left[\frac{\pi^{2}}{6}+\frac{1}{4}\left(\phi_{j}-\phi_{k}\right)\left(\phi_{j}-\phi_{k}-2\pi\right)\right].\end{eqnarray}  
The $T^{4}$ term dominates for $m/T\ll1$,
and has $P_{0}$ as a minimum provided $N_{f}>1/2$. Even if the adjoint
fermion mass is enhanced by chiral symmetry breaking, as would be
expected in a confining phase, it should be of order $gT$ or less,
and the second term in the expansion of $V_{1-loop}$ can be neglected
at sufficiently high temperture.

\section{Temporal String Tensions}

The timelike string tension $\sigma_{k}^{(t)}$ between k quarks and
k antiquarks can be measured from the behavior of the correlation
function \begin{equation}
\left<Tr_{F}P^{k}\left(\vec{x}\right)Tr_{F}P^{+k}\left(\vec{y}\right)\right>\simeq\exp\left[-\frac{\sigma_{k}^{(t)}}{T}\left|\vec{x}-\vec{y}\right|\right]\end{equation}
at sufficiently large distances. Two widely-considered scaling behaviors
for string tensions are Casimir scaling, characterized by\begin{equation}
\sigma_{k}=\sigma_{1}\frac{k\left(N-k\right)}{N-1},\end{equation}
\begin{equation}
\end{equation}
 and sine-law scaling, given by\begin{equation}
\sigma_{k}=\sigma_{1}\sin\left[\frac{\pi k}{N}\right].\end{equation}
 For a review, see reference \cite{Greensite:2003bk}.
 Timelike string tensions are calculable perturbatively
in the high-temperature confining region from small fluctuations about
the confining minimum of the effective potential \cite{Meisinger:2004pa}. 
The scale is naturally of order $gT$:
\begin{eqnarray}
\left(\frac{\sigma_{k}^{(t)}}{T}\right)^{2}
&=&g^{2}N\frac{2N_{f}m^{2}}{2\pi^{2}}\sum_{j=0}^{\infty}\left[K_{2}\left((k+jN)\beta m\right)+K_{2}\left((N-k+jN)\beta m\right)-2K_{2}\left((j+1)N\beta m\right)\right]\nonumber\\
&&-g^{2}N\frac{T^{2}}{3N^{2}}\left[3\csc^{2}\left(\frac{\pi k}{N}\right)-1\right].
\end{eqnarray}
 These string tensions are continuous functions of $\beta m$. The
$m=0$ limit is simple: \begin{equation}
\left(\frac{\sigma_{k}^{(t)}}{T}\right)^{2}=\frac{\left(2N_{f}-1\right)g^{2}T^{2}}{3N}\left[3\csc^{2}\left(\frac{\pi k}{N}\right)-1\right]\end{equation}
and is a good approximation for $\beta m\ll1$. This scaling law is
not at all like either Casimir or sine-law scaling, because the usual
hierarchy $\sigma_{k+1}^{(t)}\ge\sigma_{k}^{(t)}$ is here reversed.
Because we expect on the basis of $SU(3)$ simulations that the high-temperature
confining region is continuously connected to the conventional low-temperature
region, there must be an inversion of the string tension hierarchy
between the two regions for all $N\ge4$.

\section{Spatial String Tensions}

The confining minimum $P_{0}$ of the effective potential breaks $SU(N)$
to $U(1)^{N-1}$. This remaining unbroken Abelian gauge group naively
seems to preclude spatial confinement, in the sense of area law behavior
for spatial Wilson loops. However, as first discussed by Polyakov
in the case of an $SU(2)$ Higgs model in $2+1$ dimensions,
instantons can lead to nonperturbative confinement \cite{Polyakov:1976fu}.
In the high-temperature confining region, the dynamics of the magnetic
sector are effectively three-dimensional due to dimensional reduction.
The Polyakov loop plays a role similar to an adjoint Higgs field,
with the important difference that $P$ lies in the gauge group, while
a Higgs field would lie in the gauge algebra. The standard topological
analysis \cite{Weinberg:1979zt} is therefore slightly altered, and there are $N$ fundamental
monopoles in the finite temperature gauge theory 
\cite{Lee:1998vu,Kraan:1998kp,Lee:1998bb,Kraan:1998pm,Kraan:1998sn} with charges
proportional to the affine roots of $SU(N)$, given by
$2\pi\alpha_{j}/g$
where 
$\alpha_{j}=\hat{e}_{j}-\hat{e}_{j+1}$
for $j=1$ to $N-1$ and
$\alpha_{N}=\hat{e}_{N}-\hat{e}_{1}.$
Monopole effects will be suppressed by powers of the Boltzmann factor
$\exp\left[-E_{j}/T\right]$ where $E_{j}$ is the energy of a monopole
associated with $\alpha_{j}$. 

In the high-temperature confining region,
monopoles interact with each other through both their long-ranged
magnetic fields, and also via a three-dimensional scalar interaction, mediated by $A_{4}$.
The scalar interaction is short-ranged,
falling off with a mass of order  $gT$. The long-range properties of the magnetic sector
may be represented in a simple form by a generalized sine-Gordon model
which generates the grand canonical ensemble for the monopole/anti-monopole
gas \cite{Unsal:2008ch}. The action for this model represents the Abelian dual form
of the magnetic sector of the $U(1)^{N-1}$ gauge theory. It is given
by
\begin{equation}
S_{mag}=\int d^{3}x\left[\frac{T}{2}\left(\partial\rho\right)^{2}-2\xi\sum_{j=1}^{N}\cos\left(\frac{2\pi}{g}\alpha_{j}\cdot\rho\right)\right]\end{equation}
 where $\rho$ is the scalar field dual to the $U(1)^{N-1}$ magnetic
field. The monopole fugacity $\xi$ is given by $\exp\left[-E_{j}/T\right]$
times functional determinantal factors \cite{Zarembo:1995am}. 

This Lagrangian is a generalization of the one considered by Polyakov
for $SU(2)$, and the analysis of magnetic confinement follows along
the same lines \cite{Polyakov:1976fu}. The Lagrangian has $N$ degenerate inequivalent minima
$\rho_{0k}=g\mu_{k}$ where the $\mu_{k}$'s are the simple fundamental
weights, satisfying $\alpha_{j}\cdot\mu_{k}=\delta_{jk}$. Note that
$e^{2\pi i\mu_{k}}=z^{k}$. A spatial Wilson loop 
\begin{equation}
W\left[\mathcal{C}\right]=\mathcal{P}\exp\left[i\oint_{\mathcal{C}}dx_{j}\cdot A_{j}\right]\end{equation}
in the x-y plane
introduces a discontinuity in the z direction in the field dual to
$B$.
Moving this discontinuity out to spatial infinity, the string
tension of the spatial Wilson loop is the interfacial energy of a
one-dimensional kink interpolating between the vacua $\rho_{0k}$.
The calculation is similar to that of the 't Hooft loop in the deconfined phase,
where the kinks interpolate between the $N$ different solutions associated
with the spontaneous breaking of  $SU(N)$.
The main technical difficulty lies in finding the correct kink solutions.
A straight line ansatz through
the Lie algebra \cite{Giovannangeli:2001bh} using\begin{equation}
\rho(z)=g\mu_{k}q(z)\end{equation}
 gives\begin{equation}
\sigma_{k}^{(s)}=\frac{8}{\pi}\left[\frac{g^{2}T\xi}{N}k\left(N-k\right)\right]^{1/2}\end{equation}
This result is exact for $N=2$ or $3$, but may be only an upper bound for
$N>3$. The square-root-Casimir scaling behavior obtained
differs significantly from
both Casimir and sine-law scaling, and should be easily distinguishable
in lattice simulations.

\section{Conclusions}

We have been able to predict analytically a number of properties of
the high temperature confined region, which lattice simulations should
be able to confirm. The phase structure and thermodynamics of these
models are particularly rich for $N\ge4$. However, even in the case
of $SU(2),$ there is an interesting prediction of a first-order transition
between the deconfined phase and the high-temperature confined region
as the adjoint fermion mass is varied. It follows that there must
be a tricritical point in the $\beta-m$ plane somewhere on the critical line separating the confined
and deconfined phases. There is a perturbative prediction for
an inverted hierarchy of
timelike string tensions for $N\ge4$.
There is also a semiclassical expression
for spacelike string tensions, which are predicted to be proportional
to the square root of the monopole fugacity.

These predictions for the high-temperature region lead naturally to
additional questions that lattice simulations can address, but semiclassical
methods most likely cannot. Lattice simulations can explore the crossover
from conventional, low-temperature confining behavior to the behavior
predicted in the high-temperature confining region. Some features
can be studied in simulations where a simple deformation of the action
is used, as in \cite{Myers:2007vc}. In addition to the string tensions, these
features
include monopole and instanton densities, and the topological susceptibility.
Other aspects will require the inclusion of adjoint dynamical fermions
in lattice simulations. Chiral symmetry breaking is of particular
interest. Unsal \cite{Unsal:2007vu,Unsal:2007jx} has proposed a detailed picture of chiral symmetry
breaking which can be independently checked by simulation. 
The accessibility of lattice field configurations as well as conventional
observables makes the high-temperature confined region a natural place
to explore the overlap of theory and simulation.


\begin{thebibliography}{99}

%\cite{Gross:1980br}
\bibitem{Gross:1980br}
  D.~J.~Gross, R.~D.~Pisarski and L.~G.~Yaffe,
  \emph{QCD And Instantons At Finite Temperature},
  Rev.\ Mod.\ Phys.\  {\bf 53} (1981) 43.
  %%CITATION = RMPHA,53,43;%%

%\cite{Weiss:1980rj}
\bibitem{Weiss:1980rj}
  N.~Weiss,
  \emph{The Effective Potential For The Order Parameter Of Gauge Theories At Finite
  Temperature},
  Phys.\ Rev.\  D {\bf 24} (1981) 475.
  %%CITATION = PHRVA,D24,475;%%

%\cite{Meisinger:2001fi}
\bibitem{Meisinger:2001fi}
  P.~N.~Meisinger and M.~C.~Ogilvie,
  \emph{Complete high temperature expansions for one-loop finite temperature effects},
  Phys.\ Rev.\  D {\bf 65} (2002) 056013
  [arXiv:hep-ph/0108026].
  %%CITATION = PHRVA,D65,056013;%%

%\cite{Davies:1999uw}
\bibitem{Davies:1999uw}
  N.~M.~Davies, T.~J.~Hollowood, V.~V.~Khoze and M.~P.~Mattis,
  \emph{Gluino condensate and magnetic monopoles in supersymmetric  gluodynamics},
  Nucl.\ Phys.\  B {\bf 559} (1999) 123
  [arXiv:hep-th/9905015].
  %%CITATION = NUPHA,B559,123;%%

%\cite{Davies:2000nw}
\bibitem{Davies:2000nw}
  N.~M.~Davies, T.~J.~Hollowood and V.~V.~Khoze,
  \emph{Monopoles, affine algebras and the gluino condensate},
  J.\ Math.\ Phys.\  {\bf 44} (2003) 3640
  [arXiv:hep-th/0006011].
  %%CITATION = JMAPA,44,3640;%%

%\cite{Myers:2007vc}
\bibitem{Myers:2007vc}
  J.~C.~Myers and M.~C.~Ogilvie,
  \emph{New Phases of SU(3) and SU(4) at Finite Temperature},
  Phys.\ Rev.\  D {\bf 77} (2008) 125030
  [arXiv:0707.1869 [hep-lat]].
  %%CITATION = PHRVA,D77,125030;%%

%\cite{Myers:2008ey}
\bibitem{Myers:2008ey}
  J.~C.~Myers and M.~C.~Ogilvie,
  \emph{Exotic phases of finite temperature SU(N) gauge theories with massive
  fermions: F, Adj, A/S},
  arXiv:0809.3964 [hep-lat].
  %%CITATION = ARXIV:0809.3964;%%

%\cite{Meisinger:2001cq}
\bibitem{Meisinger:2001cq}
  P.~N.~Meisinger, T.~R.~Miller and M.~C.~Ogilvie,
  \emph{Phenomenological equations of state for the quark-gluon plasma},
  Phys.\ Rev.\  D {\bf 65} (2002) 034009
  [arXiv:hep-ph/0108009].
  %%CITATION = PHRVA,D65,034009;%%


%\cite{Greensite:2003bk}
\bibitem{Greensite:2003bk}
  J.~Greensite,
  \emph{The confinement problem in lattice gauge theory},
  Prog.\ Part.\ Nucl.\ Phys.\  {\bf 51} (2003) 1
  [arXiv:hep-lat/0301023].
  %%CITATION = PPNPD,51,1;%%
  
  %\cite{Meisinger:2004pa}
\bibitem{Meisinger:2004pa}
  P.~N.~Meisinger and M.~C.~Ogilvie,
  \emph{Polyakov loops, Z(N) symmetry, and sine-law scaling},
  Nucl.\ Phys.\ Proc.\ Suppl.\  {\bf 140} (2005) 650
  [arXiv:hep-lat/0409136].
  %%CITATION = NUPHZ,140,650;%%
  
  %\cite{Polyakov:1976fu}
\bibitem{Polyakov:1976fu}
  A.~M.~Polyakov,
  \emph{Quark Confinement And Topology Of Gauge Groups},
  Nucl.\ Phys.\  B {\bf 120} (1977) 429.
  %%CITATION = NUPHA,B120,429;%%
  
  %\cite{Weinberg:1979zt}
\bibitem{Weinberg:1979zt}
  E.~J.~Weinberg,
  \emph{Fundamental Monopoles And Multi-Monopole Solutions For Arbitrary Simple
  Gauge Groups},
  Nucl.\ Phys.\  B {\bf 167} (1980) 500.
  %%CITATION = NUPHA,B167,500;%%
  
  %\cite{Lee:1998vu}
\bibitem{Lee:1998vu}
  K.~M.~Lee,
  \emph{Instantons and magnetic monopoles on R**3 x S(1) with arbitrary simple
  gauge groups},
  Phys.\ Lett.\  B {\bf 426} (1998) 323
  [arXiv:hep-th/9802012].
  %%CITATION = PHLTA,B426,323;%%
  
  %\cite{Kraan:1998kp}
\bibitem{Kraan:1998kp}
  T.~C.~Kraan and P.~van Baal,
  \emph{Exact T-duality between calorons and Taub - NUT spaces},
  Phys.\ Lett.\  B {\bf 428} (1998) 268
  [arXiv:hep-th/9802049].
  %%CITATION = PHLTA,B428,268;%%
  
  %\cite{Lee:1998bb}
\bibitem{Lee:1998bb}
  K.~M.~Lee and C.~h.~Lu,
  \emph{SU(2) calorons and magnetic monopoles},
  Phys.\ Rev.\  D {\bf 58} (1998) 025011
  [arXiv:hep-th/9802108].
  %%CITATION = PHRVA,D58,025011;%%
  
  %\cite{Kraan:1998pm}
\bibitem{Kraan:1998pm}
  T.~C.~Kraan and P.~van Baal,
  \emph{Periodic instantons with non-trivial holonomy},
  Nucl.\ Phys.\  B {\bf 533} (1998) 627
  [arXiv:hep-th/9805168].
  %%CITATION = NUPHA,B533,627;%%
  
  %\cite{Kraan:1998sn}
\bibitem{Kraan:1998sn}
  T.~C.~Kraan and P.~van Baal,
  \emph{Monopole constituents inside SU(n) calorons},
  Phys.\ Lett.\  B {\bf 435} (1998) 389
  [arXiv:hep-th/9806034].
  %%CITATION = PHLTA,B435,389;%%
  
  %\cite{Unsal:2008ch}
\bibitem{Unsal:2008ch}
  M.~Unsal and L.~G.~Yaffe,
  \emph{Center-stabilized Yang-Mills theory: confinement and large $N$ volume
  independence},
  arXiv:0803.0344 [hep-th].
  %%CITATION = ARXIV:0803.0344;%%
  
  %\cite{Zarembo:1995am}
\bibitem{Zarembo:1995am}
  K.~Zarembo,
  \emph{Monopole determinant in Yang--Mills theory at finite temperature},
  Nucl.\ Phys.\  B {\bf 463} (1996) 73
  [arXiv:hep-th/9510031].
  %%CITATION = NUPHA,B463,73;%%
  
 %\cite{Giovannangeli:2001bh}
\bibitem{Giovannangeli:2001bh}
  P.~Giovannangeli and C.~P.~Korthals Altes,
  \emph{'t Hooft and Wilson loop ratios in the QCD plasma},
  Nucl.\ Phys.\  B {\bf 608} (2001) 203
  [arXiv:hep-ph/0102022].
  %%CITATION = NUPHA,B608,203;%%
      
  %\cite{Unsal:2007vu}
\bibitem{Unsal:2007vu}
  M.~Unsal,
  \emph{Abelian duality, confinement, and chiral symmetry breaking in QCD(adj)},
  Phys.\ Rev.\ Lett.\  {\bf 100} (2008) 032005
  [arXiv:0708.1772 [hep-th]].
  %%CITATION = PRLTA,100,032005;%%
  
  %\cite{Unsal:2007jx}
\bibitem{Unsal:2007jx}
  M.~Unsal,
  \emph{Magnetic bion condensation: A new mechanism of confinement and mass gap in
  four dimensions},
  arXiv:0709.3269 [hep-th].
  %%CITATION = ARXIV:0709.3269;%%
  

\end{thebibliography}
\end{document}